\documentclass[11pt,twoside]{article}
\usepackage{asp2010}
\usepackage{subfigure}
\usepackage{natbib}

\resetcounters

\begin{document}

\title{Application of GPUs for the calculation of two point correlation functions in cosmology}

\author{Rafael~Ponce$^1$, Miguel~C\'ardenas-Montes$^1$, Juan Jos\'e Rodr\'iguez-V\'azquez$^1$, Eusebio S\'anchez$^1$ and Ignacio Sevilla$^1$}
\affil{$^1$Centro de Investigaciones Energ\'eticas Medioambientales y Tecnol\'ogicas, Av.Complutense 40, 28040 Madrid, Spain}

\begin{abstract}
In this work, we have explored the advantages and
drawbacks of using GPUs instead of CPUs in the calculation
of a standard 2-point correlation function algorithm,
which is useful for the analysis of Large Scale Structure of
galaxies.  Taking into account the huge volume of data
foreseen in upcoming surveys, our main goal has been to
accelerate significantly the analysis codes.We find that
GPUs offer a 100-fold increase in speed with respect to a
single CPU without a significant deviation in the
results. For comparison's sake, an MPI version was developed
as well. Some issues, like code implementation, which arise
from using this option are discussed.
\end{abstract}

\section{Introduction}
The two-point correlation function (2pcf) is a simple
statistic that quantifies the clustering of a given
distribution of objects. In studies of the Large Scale
Structure (LSS) of the Universe, this is an important tool
containing information about the matter clustering and the
Universe evolution at different cosmological epochs,~\cite{peebles.1980}. A classical application of this statistic is the
galaxy-galaxy correlation function to find constraints on
the matter density parameter $\Omega_m$,~\cite{MNR:MNR7063}, or the
location of the baryonic acoustic oscillation peak,~\cite{2011MNRAS.411..277S}. Other examples include cross-correlation
of background galaxies with the shear of objects caused by
the gravitational effect on light (weak lensing),~\cite{PhysRevD.78.043508}.

The 2pcf measures the excess probability of finding a couple
of galaxies separated by spatial distance $r$ or angular
distance $\theta$ with respect to the probability of finding
a couple of galaxies separated by the same distance or angle
in a random and uniform distribution. In this work we have
used the angular version of the correlation function
$w(\theta)$ though results are extendible to the
3-dimensional variant as well.

Landy \& Szalay,~\cite{1993ApJ...412...64L}, found an estimator with minimum
variance which is the standard one used in cosmological analyses:

\begin{equation}
\begin{array}{l}
   \omega(\theta) = 1 \;+\; (\frac{N_{random}}{N_{real}})2 \cdot
\frac{DD(\theta)}{RR(\theta)} \;-\; 2 \cdot
(\frac{N_{random}}{N_{real}}) \cdot \frac{DR(\theta)}{RR(\theta)}
\end{array}
  \label{eq:TPCF}
\end{equation}

where $N_{gal}$ is the number of galaxies in a real catalog,
$N_{rd}$ is the number of galaxies in a random catalog, $DD(\theta)$
is the number of pairs separated by an angular distance
$\theta$ in the real catalog, $RR(\theta)$ is the number of
pairs separated by an angular distance $\theta$ in the random
catalog and $DR(\theta)$ is the number of pairs separated by
an angular distance $\theta$ in the real catalog with respect
to the random catalog.

\section{Computational problem and previous work}
The calculation of 2pcf, Eq.\ref{eq:TPCF},
is very costly computationally so alternative strategies
have been designed to approach the problem (pixelization of the
map,~\cite{0067-0049-151-1-1}, k-trees,~\cite{Moore:astro-ph0012333}), usually at the cost of some loss
of information.

Alternatively, in~\cite{Roeh:2009:ACD:1513895.1513896}, this problem has been treated with
GPUs using a different strategy in terms of shared memory
usage. In particular, the authors of~\cite{Roeh:2009:ACD:1513895.1513896} have used a
'chessboard' strategy where arrays are passed to the global
memory. This has the disadvantage of having
restrictions in the input sample. Also, the particular
implementation in~\cite{Roeh:2009:ACD:1513895.1513896} obtained results in $cos\theta$ space, thus complicating the cosmological
interpretation of the result.

\section{Implementation and hardware}
We have implemented in CUDA the Landy-Szalay estimator with the following key features:
\begin{itemize}
\item Usage of shared memory (instead of global memory) for the dot
product and arc-cosine operations necessary to extract the
angle between two objects.
\item Application of atomic operations in shared memory to make use
efficiently of multi-threading when filling up the histograms
(DD, DR and RR in Eq.~\ref{eq:TPCF}). Partial histograms are generated in parallel in
shared memory and later combined in a single histogram, in global memory.
\item In one of the architectures we had available, we
applied a multi-GPU solution using 3 GPUs, one for each of
the histograms, in which DD and RR where used in one of the
boards containing 2 GPUs and DR in the other for maximum
efficiency.
\end{itemize}

A full description of the algorithm and its implementation can be found in~\cite{Cardenas_DES_unpublished}.
The hardware we have used to test our codes is in Table \ref{tab:hard}.

\begin{table}[!ht]
\begin{center}
\begin{tabular}{ | c| c | c | }
\hline
CPU & GPU & MPI \\ \hline 
CPU with two Intel & GTX295 & 1920 cores (two \\ 
Xeon E5520 processors & C1060 (Tesla) & Intel Xeon E5570 \\
at 2.27 GHz & C2050 (Tesla) & at 2.93 GHz, per node) \\ \hline
\end{tabular}
\caption{Hardware specifications that we have used.}
\label{tab:hard}
\end{center}
\end{table}

\section{Results and analysis}
The galaxy catalogs used are publicly available from the MICE project,~\cite{2008MNRAS.391..435F,2010MNRAS.403..1353F}.

In Table \ref{tab:gpu_cpu} we present a comparison between the execution time of CPU implementation and the execution time of GPU implementation.

\begin{table}[!ht]
\begin{center}
\begin{tabular}{ | c | c | c | c | c | }
\hline
Input file lines & CPU (s) & GTX295 (s) & C1060 (s) & C2050 (s) \\ \hline 
     $0.43\cdot10^6$ & $3.60\cdot10^4$ & $3.01\cdot10^2$ & $2.91\cdot10^2$ & $2.19\cdot10^2$ \\ \hline
     $0.86\cdot10^6$ & $1.44\cdot10^5$ & $1.20\cdot10^3$ & $1.16\cdot10^3$ & $8.76\cdot10^2$ \\ \hline
     $1.00\cdot10^6$ & $1.98\cdot10^5$ & $1.61\cdot10^3$ & $1.56\cdot10^3$ & $1.17\cdot10^3$ \\ \hline
     $1.29\cdot10^6$ & $3.24\cdot10^5$ & $2.68\cdot10^3$ & $2.59\cdot10^3$ & $1.97\cdot10^3$ \\ \hline
     $1.72\cdot10^6$ & $5.76\cdot10^5$ & --------- & $4.64\cdot10^3$ & $3.51\cdot10^3$ \\ \hline
     $3.45\cdot10^6$ & $2.32\cdot10^6$ & --------- & $1.88\cdot10^4$ & $1.41\cdot10^4$ \\ \hline
     $6.89\cdot10^6$ & $9.22\cdot10^6$ & --------- & $7.45\cdot10^4$ & $5.61\cdot10^4$ \\ \hline
   \end{tabular}
\caption{Comparison between CPU execution time and diverse GPUs execution time.}
\label{tab:gpu_cpu}
\end{center}
\end{table}

\begin{figure}[!ht]
\centering
\subfigure{\includegraphics[width=0.48
\textwidth,angle=0]{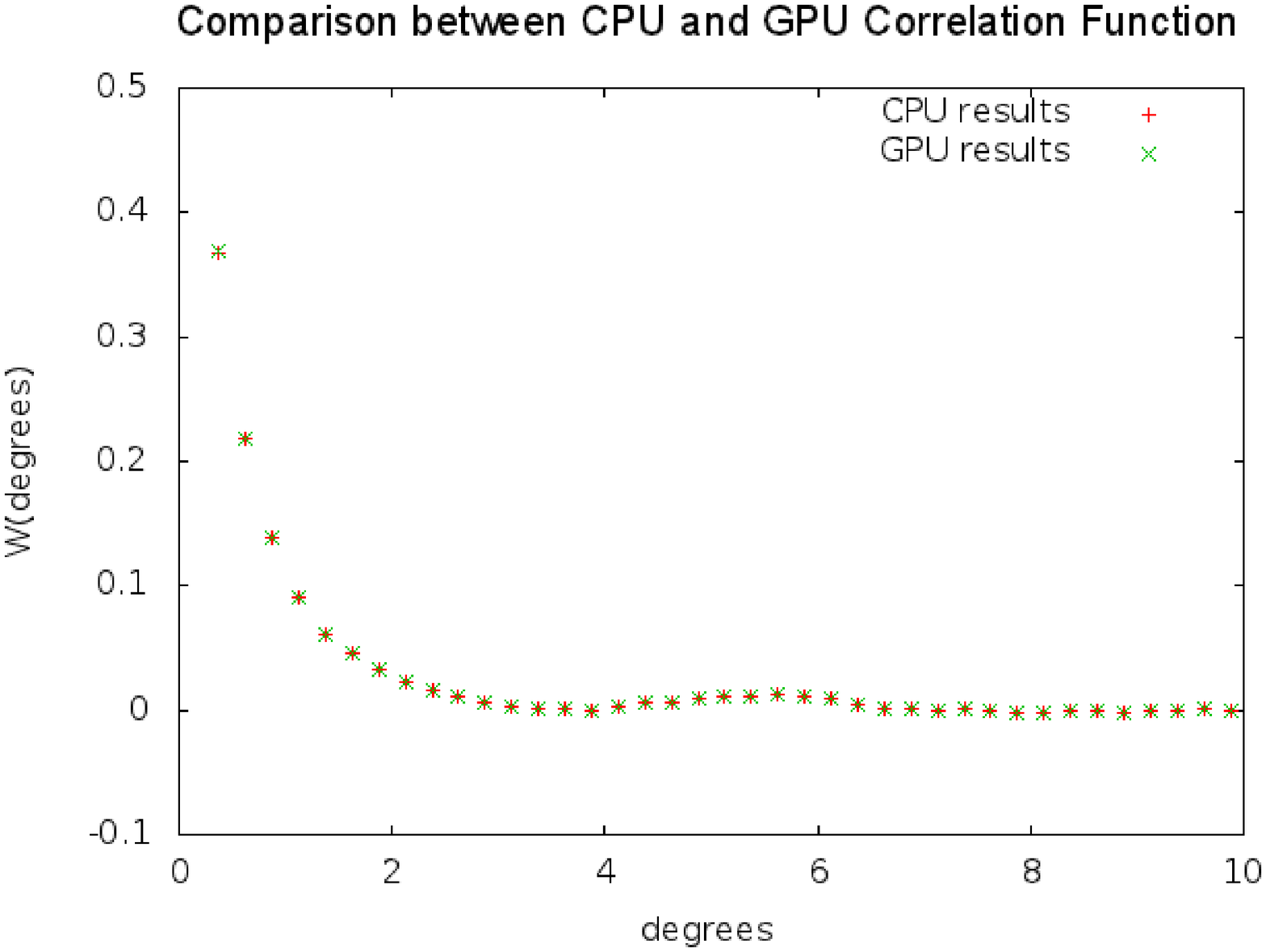}\label{fig:results_wtheta}}
\subfigure{\includegraphics[width=0.48
\textwidth,angle=0]{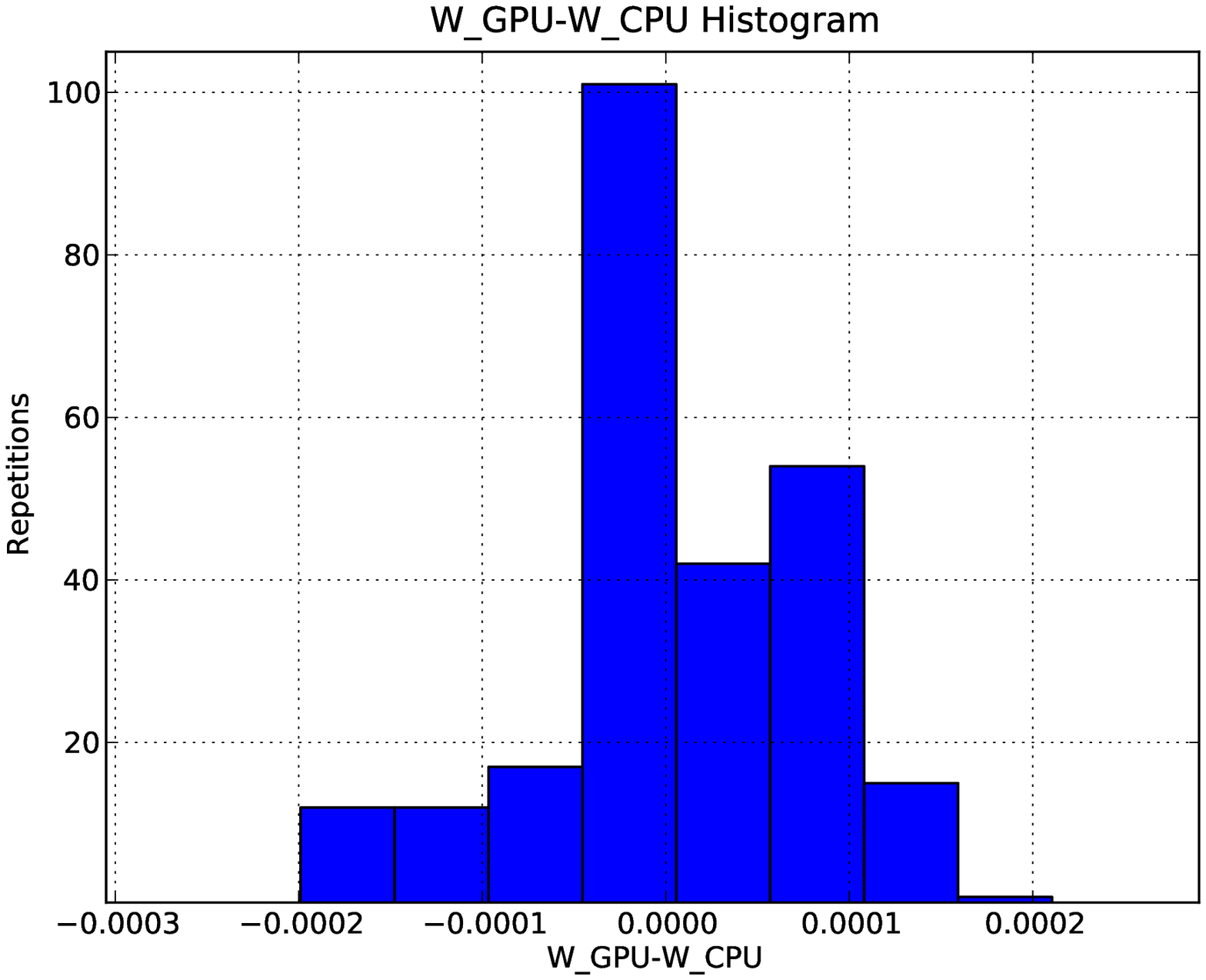}\label{fig:results_residuals}}
\caption{Panel (a, left) shows a comparison between correlation functions, the red one was calculated with the CPU code, the green one with the GPU code, while panel (b, right) shows
the residuals between GPU and CPU codes. These residuals are really small and fall into the statistical errors.}
\end{figure}

In Fig.~\ref{fig:results_wtheta} we show, for MICE catalog, one of the
correlation functions calculated using this code, versus the
same calculation using a standard implementation in C for
CPUs, for reference. The residuals at each point are
plotted in Fig.~\ref{fig:results_residuals} and are far below the
expected errors due to cosmic variance, i.e., the
statistical errors due to the small number of 'fields'
available in the sky.

\begin{figure}[!ht]
\centering
\subfigure{\includegraphics[width=3.8in,height=2.5in]{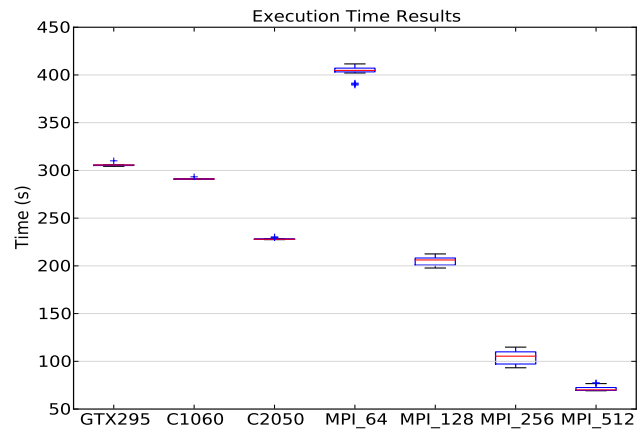}}
\caption{GPU and MPI execution time results.}\label{fig:boxplot}
\end{figure}

We have also done a comparison between GPUs and MPI. In
Fig.~\ref{fig:boxplot} we have our MPI time with GPUs time like a boxplot graphic.

\section{Conclusions}
We have developed an implementation of the Landy-Szalay
two-point correlation function in CUDA to make use of the
power GPUs have to offer in terms of parallelization. The
speed-up with respect to a CPU is 164-fold (C2050) using the
same algorithm. With respect to an implementation of k-trees
in CPUs we obtain an increase of 6.4-fold (for
$0.43\cdot10^6$ objects). Several MPI configurations have
been explored being the GPU implementation surpassed by the
usage of more than 64 nodes, see Fig.~\ref{fig:boxplot}.

Some options to be explored remain, such as full-blown
multi-GPU implementation, coding the k-trees or extending
the work to higher order correlation functions, for other
types of cosmological analyses such as understanding
non-Gaussianities in the primordial perturbations. 

\acknowledgements We would like to the thank ASP for the chance to present at ADASS and we also acknowledge the use of data from the MICE simulations, publicly available at http://www.ice.cat/mice. 

\bibliography{P121}
\end{document}